\documentclass[fleqn,usenatbib]{mnras}

\usepackage{newtxtext,newtxmath}

\usepackage[T1]{fontenc}
\usepackage{ae,aecompl}
\usepackage{float}
\usepackage{lastpage}
\usepackage{enumerate}
\usepackage{fancyhdr}
\usepackage{mathrsfs}
\usepackage{xcolor}
\usepackage{listings}
\usepackage{hyperref}
\usepackage{lscape}
\usepackage{rotating}

\usepackage{graphicx}	
\usepackage{amsmath}	
\usepackage[justification=centering]{caption}
\usepackage{multirow}

\title[$^{12}$C($\alpha$, $\gamma$)$^{16}$O reaction rate and stellar evolution]{The impact of the  uncertainties in the $^{12}$C($\alpha$, $\gamma$)$^{16}$O reaction rate on the evolution of low-- to intermediate--mass stars}

\author[B. T. Pepper et al.]{Ben T. Pepper,$^{1}$\thanks{E-mail: ben.pepper2012@gmail.com}
A. G. Istrate$^{2}$,
A. D. Romero$^{1}$ and 
S. O. Kepler$^{1}$
\\
$^{1}$Physics Institute, Universidade Federal do Rio Grande do Sul, 91501-900 Porto-Alegre, RS, Brazil\\
$^{2}$Department of Astrophysics/IMAPP, Radboud University, P O Box 9010, NL-6500 GL Nijmegen, The Netherlands\\
}

\date{Accepted XXX. Received YYY; in original form ZZZ}

\pubyear{2021}

\begin{document}
\label{firstpage}
\pagerange{\pageref{firstpage}--\pageref{lastpage}}
\maketitle

\begin{abstract}

One of the largest uncertainties in stellar evolutionary computations is the accuracy of the considered reaction rates. The $^{12}$C($\alpha$,~$\gamma$)$^{16}$O reaction is particularly important for the study of low- and intermediate-mass stars as it determines the final C/O ratio in the core which influences the white dwarf cooling evolution. Thus, there is a need for a study of how the computations of white dwarfs and their progenitors that are made to date may be affected by the uncertainties of the $^{12}$C($\alpha$,~$\gamma$)$^{16}$O reaction rates. In this work we compute fully evolutionary sequences using the \textsc{MESA} code with initial masses in the range of $0.90 \leq M_i/M_\odot \leq 3.05$. We consider different adopted reaction rates, obtained from the literature, as well as the extreme limits within their uncertainties.  As expected, we find that previous to the core helium burning stage, there are no changes to the evolution of the stars. However, the subsequent stages are all affected by the uncertainties of the considered reaction rate. In particular, we find differences to the convective core mass during the core helium burning stage which may affect pulsation properties of subdwarfs, the number of thermal pulses during the asymptotic giant branch and trends between final oxygen abundance in the core and the progenitor masses of the remnant white dwarfs.

\end{abstract}




\section{Introduction} \label{intro}


Single stellar evolution is fuelled by nuclear reactions that occur within the stellar interior \citep{bethe1939, hoyle46, hoyle54, burbidge57}. These reactions not only release energy which allows the star to support itself against gravitational collapse and remain in hydrostatic equilibrium, but also change the composition of the star: this is known as nucleosynthesis \citep{eddington1920, hoyle54, burbidge57}. The study of these nuclear reactions is where nuclear physics and astronomy come hand-in-hand; an understanding of what happens at the fundamental level provides a better knowledge of how stars evolve and influence their environment. Particularly, improved estimations of the often uncertain reaction rate data, including formula fitted to such data, will improve the accuracy of stellar evolution codes and the understanding of stellar evolution \citep{CF88,angulo99, katsuma12, xu13, an16}. Such estimations are hereafter referred to as 'reaction rates'.

The $^{12}$C($\alpha$, $\gamma$)$^{16}$O reaction during the central helium burning stage is considered to be the most important mechanism for defining the white dwarf (WD) core composition \citep{salaris05, dantona90, DG17, deboer19}. However, the reaction rate for this reaction has an extremely large uncertainty \citep{fowler67, CF88, kunz02, an16, deboer17, deboer19}.
The main entrance channel for the $^{12}\text{C} + \alpha$ mechanism ($E_{\alpha_0} = 7.16\, \text{MeV}$) does not have a resonance channel close to this threshold, the closest occurring at $E_x = 9.59\, \text{MeV}$. Instead, the low energy cross-section is largely influenced by the $1^{-1}$ ($E_x = 7.12\, \text{MeV}$) and $2^+$ ($E_x = 6.92\, \text{MeV}$) subthreshold states \citep[see Figure 2 of][for details]{deboer17}. The primary influence of these two nearby subthreshold states and the addition of possible resonant transitions in the wings of the broad channel at $E_x = 9.59\, \text{MeV}$ makes the nuclear cross-section extremely difficult to estimate \citep[see][]{fowler67, kunz02, an16, deboer17, deboer19, aliotta21}. 


During the core helium burning (CHB) stage, carbon is produced from the fusion of three helium nuclei via the triple-$\alpha$ process \citep{salpeter52b, kipp90, salaris05, prialnik09}. As the abundance of helium in the core depletes, the probability of carbon interacting with helium to produce oxygen [via $^{12}$C($\alpha$, $\gamma$)$^{16}$O] is larger than that of the triple-$\alpha$ process at late times during the core helium burning stage \citep{salaris05}. Thus, the $^{12}$C($\alpha$, $\gamma$)$^{16}$O reaction is of great importance and is vital to model the carbon-oxygen (C/O) abundance in the inner chemical profiles for all stellar masses, but particularly low-- and intermediate--mass stars  \citep{woosley95,weaver93, wallerstein97}. 

The C/O abundance, therefore the $^{12}$C($\alpha$, $\gamma$)$^{16}$O reaction, is important in many areas of stellar evolution. Such as, influencing the pulsation properties of ZZ Ceti stars \citep{DG15, DG17}. Differences between the considered $^{12}$C($\alpha$, $\gamma$)$^{16}$O reaction rate will also affect the duration of the core helium burning stage \citep[][]{deboer17}. In addition, the $^{12}$C($\alpha$, $\gamma$)$^{16}$O reaction impacts supernova explosions as the outcome is related to the composition of the final WD \citep[e.g.][]{iben84, wu20} and third dredge-up episodes (TDUs) during the asymptotic giant branch (AGB) stage \citep{frost96, karakas02, marigo02, karakas03, marigo07, cristallo09, weiss09, ventura09, kalirai14, matteucci21}. Furthermore, thermonuclear explosions of C/O WDs impacts the ignition of Type 1a supernovae, an important event in constraining cosmological parameters \citep{perlmutter99, riess98}.
The enrichment of the outer layer of the AGB stars from dredge-up and the mass-loss affects the chemical evolution of galaxies \citep{matteuccibook12, boothroyd88, kobayashi20, ventura20, cristallo15, matteucci21}. Additionally, the $^{12}$C($\alpha$, $\gamma$)$^{16}$O reaction governs whether a star will form a neutron star or black hole \citep{brown01, heger02, tur07, west13, sukhbold20}. Gravitational wave detections from black hole mergers can also be used to constrain the $^{12}$C($\alpha$, $\gamma$)$^{16}$O reaction rate by determining the mass of the black hole and the fraction of carbon and oxygen that remains \citep[see][for details]{farmer20}.

\citet{DG15} and \citet{DG17} consider 3 different reaction rates: an adopted rate from \citet{angulo99} and the high and low rates from \citet{kunz02}. They consider these alternate rates for the CHB until the thermally pulsing asymptotic giant branch (TP-AGB) phase with a sole focus on how the pulsational properties are affected in ZZ Ceti stars, rather than all stages as we attempt in this work.




In this work, we use stellar evolutionary models as tools to study the impact of the $^{12}$C($\alpha$, $\gamma$)$^{16}$O reaction rate uncertainties on the stellar structure and evolution of low-- and intermediate--mass stars. The paper is organised as follows. Section~\ref{CT} describes the input physics and numerical tool used to compute the evolutionary sequences, as well as a deeper discussion of the considered $^{12}$C($\alpha$, $\gamma$)$^{16}$O reaction rates used in this work. In section~\ref{resdis} we present and discuss our results. We summarise our work in section~\ref{conc}, concluding our findings and indicating future areas where the impact of this work may affect.

\section{Numerical Tools} \label{CT}

\subsection{MESA Input Physics}


In this work we employ the Modules for Experiments in Stellar Astrophysics (MESA) code version-r15140 (see \citet{paxton11,paxton13,paxton15,paxton18}, for details). We compute the full evolutionary sequence from the zero age main sequence (ZAMS) through both core hydrogen and helium burning stages, leading to the AGB and the white dwarf stage (WD). The computation stops when the stellar model reaches a luminosity of $\log(L/L_\odot) = -3$ on the WD cooling track. 
This stopping condition is applied such that the sequences have experienced their evolution through the DAV instability strip \citep{fontaine08, winget08, althaus10}. This allows for asteroseismology of ZZ Ceti stars to be performed in the future. The final WD masses obtained in this work range from $0.513M_\odot \leq M_f / M_\odot \leq 0.691\,M_\odot$. The initial mass range considered in this work is selected such that all sequences evolve into a carbon--oxygen WD \cite[examples of works which consider/include a similar mass range are][]{renedo10,romero15,DG17,marigo20}. 



We compute a total of 246 sequences, with an initial metallicity of $Z_i = 0.01$ and 41 initial masses in the range of $0.90 \leq M_i/M_\odot \leq 3.05$. For each initial mass, we compute the full evolution considering 6 different formulae for the $^{12}$C($\alpha$, $\gamma$)$^{16}$O reaction rate. The 6 reaction rates are adapted from \citet{angulo99} and \citet{an16}. Each source comprises 3 reaction rates: the adopted rate, the low and high limiting values, given by the reported uncertainties of the respective rate (see Section~\ref{c12o16_rate}). The rates taken from \citet{angulo99} are part of the NACRE compilation and have been used extensively in other computations \citep{renedo10, romero15, DG17}. The reaction rates from \citet{an16} are less recognised, but boast a lower uncertainty on their reported adopted reaction rate. More detail on these rates and their significance can be found in Section~\ref{c12o16_rate}.



We use the reaction network 'basic.net', which comprises 33 individual reactions including the full p-p chain, CNO cycle, 3$\alpha$ up until $^{24}$Mg, which contains the $^{12}$C($\alpha$, $\gamma$)$^{16}$O reaction. This network also includes 8 individual isotopes: $^1$H, $^3$He, $^4$He, $^{12}$C, $^{14}$N, $^{16}$O, $^{20}$Ne, $^{24}$Mg in addition to elementary and $\alpha$ particles.



In our computations, we consider the default radiative opacity tables within \textsc{MESA}. These are from \citet{ferguson05} (for $2.7 \leq \textrm{log}T \leq 4.5$) and from the OPAL project (for $3.75 \leq \textrm{log}T \leq 8.7$) \citep{iglesias93, iglesias96}. Furthermore, we consider OPAL Type~2 tables as they allow for varying amounts of C and O, which are needed for helium burning and beyond \citep{iglesias96, paxton11}.



We adopt the standard mixing length free parameter as $\alpha = 2.0118$. This value is adopted from the work of \citet{guzik16} who found this value to be a good approximation for sequences that consider the solar metallicity when using the opacity tables from the OPAL project. To derive this value, \citet{guzik16} compared calculated non-adiabatic solar oscillation frequencies and solar interior sound speeds to observed frequencies and helioseismic inferences. However, it should be noted that \citet{guzik16} consider an initial metallicity of $Z_i = 0.015$, rather than the value we consider in this work ($Z_i = 0.01$). Such a difference would alter the value of the $\alpha$ parameter if a similar analysis was performed with this initial metallicity consideration. Convective mixing is treated as a time-dependent diffusion process, with the diffusion coefficient given as,
\begin{equation} \label{eq: D_ov}
    D_\mathrm{EM} = D_0 \exp(-2z/fH_P)
\end{equation}
\noindent where $H_P$ is the pressure scale-height at the convective boundary, $D_0$ is the diffusion coefficient of the unstable regions that are near the convective boundary, and $z$ is the geometric distance from the convective boundary. $f$ is an adjustable free parameter that controls the efficiency of mixing by setting the size of the overshooting region \citep{herwig97, herwig2000}. We take the value of $f = 0.016$ for all regions of the model for this work, following the same consideration of overshooting as \citet{herwig2000, weiss09, DG17}. This treatment of the convective boundaries was also adopted by other authors for single stellar evolution computations \citep{weiss09, romero15, DG17}.

The presence of dredge-up episodes during the core helium burning stage is relevant for the final composition of WDs \citep{prada02, straniero03, renedo10}. During the thermally pulsing AGB phase, although overshooting was considered at the boundary of the convective H-rich envelope during the TP-AGB, the third dredge-up episodes did not occur. Therefore, the evolution of the hydrogen--exhausted core (which is hereafter simply referred to as "the helium core mass") and the final mass of the sequences for those which should experience some third dredge-up episodes will be affected (see Section~\ref{AGB}). We define the "helium core mass" as the region from the centre until the local abundance of hydrogen is greater than $10^{-6}$.
Additional models were computed to assess the impact of the third dredge-up on the core mass growth during the thermal pulses (see Section~\ref{AGB} and Appendix~\ref{model_appendix} for details). 

For regions stable against convection according to the Ledoux criterion, but there is an inversion of mean molecular weight, we employ thermohaline mixing. In \textsc{MESA} this is treated as a diffusion process, as above, with a diffusion coefficient produced by the stability analysis of \citet{ulrich72} and \citet{kipp80}. For the efficiency parameter of thermohaline mixing, we consider $\alpha_{th} = 1.0$ \citep[see Equation~14 of][for details]{paxton13}).
Thermohaline mixing was considered in order to smooth a discontinuity in the carbon and oxygen chemical profiles at the edge of the C/O core, during the early-AGB.



Towards the end of the core helium burning stage, when the central He abundance is lower than $\sim$10\%, breathing pulse--like instabilities may appear. However, these events are attributed to adopted algorithms rather than to the physics of convection \citep[see][for details]{straniero03, romero15, constantino16, constantino17}. To suppress the breathing pulses, when the central abundance of He drops below 0.13, we neglect convection until the central abundance of helium decreases below $10^{-6}$, similar to the prescription used by \citet{renedo10} and \citet{romero15}. Without this prescription, the final carbon-to-oxygen (C/O) ratios can vary rapidly (up to $\pm 0.1$) with small increments of initial mass ($0.05\,M_\odot$).



During the main sequence (MS), red--giant branch (RGB) and core helium burning stages, the mass-loss due to stellar winds follows the rate based on the Reimers formula \citep[see][]{reimers75}. The asymptotic giant branch and subsequent evolution follow a rate based on the Bloecker formula instead \citep[see][]{bloecker95}. We set our scale factors to be $\eta_R = 0.5$ and $\eta_B = 0.2$ for the Reimers and Bloecker formulae, respectively. These values are chosen as they reproduce a WD with a similar final mass to that found by \citet{renedo10} for $M_i = 1.00M_\odot$ with $Z_i = 0.01$.



A grey atmosphere is employed for the entire evolution of all sequences, which utilises the grey Eddington $\tau$ relation. We consider the equations of state ELM EOS and DT2 EOS, which are derived from the HELM EOS \citep{timmes00} and the SCVH tables \citep{saumon95}, respectively.

Once the star leaves the AGB, we employ an element diffusion process from the work of \citet{burgers69}. We refer to element diffusion as the physical mechanism for mixing chemicals that includes gravitational settling, thermal diffusion and chemical diffusion. Gravitational settling leads to denser element diffusing towards the core, while lighter elements float towards the surface. Thermal diffusion acts in the same direction as gravitational settling, although to a lesser extent, bringing highly charged and more massive species to the central regions of the star. Chemical diffusion, however, works against this general direction \citep[see][for details]{iben85, thoul94}. In addition to the aforementioned processes, \textsc{MESA} includes radiative accelerations \citep{hu11} into their element diffusion prescription. These radiative forces are negligible in hot regions, as well as being computationally demanding. Hence, we do not consider the effects of radiative levitation. Our element diffusion process is applied to the following isotopes: $^1$H, $^3$He, $^4$He, $^{12}$C, $^{14}$N, $^{16}$O, $^{20}$Ne, $^{24}$Mg.




\subsection{The $^{12}$C($\alpha$, $\gamma$)$^{16}$O Reaction} \label{c12o16_rate}

Here we discuss a brief, yet relevant, history of $^{12}$C($\alpha$, $\gamma$)$^{16}$O reaction rate evaluations. We lead this into further detail for the $^{12}$C($\alpha$, $\gamma$)$^{16}$O reaction rate prescriptions from \citet{angulo99} and \citet{an16}, discussing their differences to the previous determinations from the literature.



\citet{fowler67} organised the first symposium of reaction rate cross-sections that included the $^{12}$C($\alpha$, $\gamma$)$^{16}$O reaction. At the time, many resonant factors were neglected and were updated by \citet{CF88}. However, it is believed that some resonances were still neglected and the treatment of the S-factor in this work produced values that are too small and require a scale factor of $\sim$2 to produce a realistic S-Factor \citep{angulo99,  kunz02, heil08, an16, deboer17, deboer19}.


Built upon the works of \citet{fowler67, CF88} and those associated works in between, \citet{angulo99} provided a strong basis for the $^{12}$C($\alpha$,~$\gamma$)$^{16}$O reaction within the NACRE compilation. \citet{angulo99} provided the reaction rates for 86 different reactions, including $^{12}$C($\alpha$, $\gamma$)$^{16}$O. For the S-factor calculations, \citet{angulo99} considered the values for non-resonant energies. For narrow resonances, however, they fit the resulting cross-section using a Briet-Wigner model. When the effects of different resonant energies overlap, they use a multi-resonance fit, shown in equation 29 of \citet{angulo99}. \citet{angulo99} state that their analysis is numerical for the majority, although they do provide an analytical approach for each reaction, for completeness. They find that their numerical approach yields a higher accuracy for their calculated reaction rates. The quoted S-Factor value from \citet{angulo99} for a stellar energy of $\text{300\,keV}$ is $S(\text{300\, \text{keV}}) =  199 \pm 64 \, \text{keV\,b}$, resulting in a reaction rate ($RR$) of $RR(\text{300\, \text{keV}}) = (9.11^{+3.69}_{-3.67}) \cdot 10^{-15} \text{cm\textsuperscript{3}\, mole\textsuperscript{-1} s\textsuperscript{-1}}$. A stellar energy of $E = \text{300\, \text{keV}}$ is often chosen as the energy at which to compare the S-factors across different works, as it is associated with the ignition of core helium burning. In this work, we consider the adopted rate of \citet{angulo99} (NACRE\_A) and the highest and lowest reaction rate within the uncertainties (NACRE\_H and NACRE\_L, respectively). Hereafter, we refer to the collective $^{12}$C($\alpha$, $\gamma$)$^{16}$O reaction rates from \citet{angulo99} as 'NACRE'.



\citet{an16} point out that the resonance parameters used by \citet{kunz02}, which were taken from \citet{tilley93}, neglect the ground state transitions from the works of \citet{brochard73, ophel76}. This results in a larger value for the expected reaction rate at helium burning temperatures. Instead, \citet{an16} use the reduced R-matrix and S-factor derived by \citet{an15} to estimate the reaction rate, which accounted for all transitions. 


In their computations, \citet{an15} and \citet{an16} found a significant reduction to the uncertainty of their S-factors when compared to that of \citet{angulo99},  $S(\text{300\, keV}) =  162.7 \pm 7.3 \, \text{keV\, b}$. The reaction rate for the same energy resulted $RR(\text{300\,keV}) =  (7.83 \pm 0.35) \times 10^{-15} \text{cm\textsuperscript{3}\, mole\textsuperscript{-1} s\textsuperscript{-1}}$. We consider the adopted rate from \citet{an16} (An\_A) and the highest and lowest reaction rate within the uncertainties (An\_H and An\_L, respectively). However, the S-factor calculation of \citet{an15}, seems to neglect external contributions for ground state energy levels, making this approximation not valid for high precision analysis \citep{deboer17}. Therefore, we treat the uncertainties of the $^{12}$C($\alpha$, $\gamma$)$^{16}$O reaction rate from \citet{an16} as arbitrary differences to determine the effect of the urgent need for more precise $^{12}$C($\alpha$, $\gamma$)$^{16}$O reaction rate uncertainties, as claimed by \citet{kunz02, tur10}. Some works further claim that the uncertainty must be less than 10\% to be on par with non-nuclear physical uncertainties \citep[see][for details]{woosley03, deboer17}.



\begin{figure}
    \centering
    \includegraphics[trim = 5mm 0mm 0mm 0mm, width = 1\linewidth]{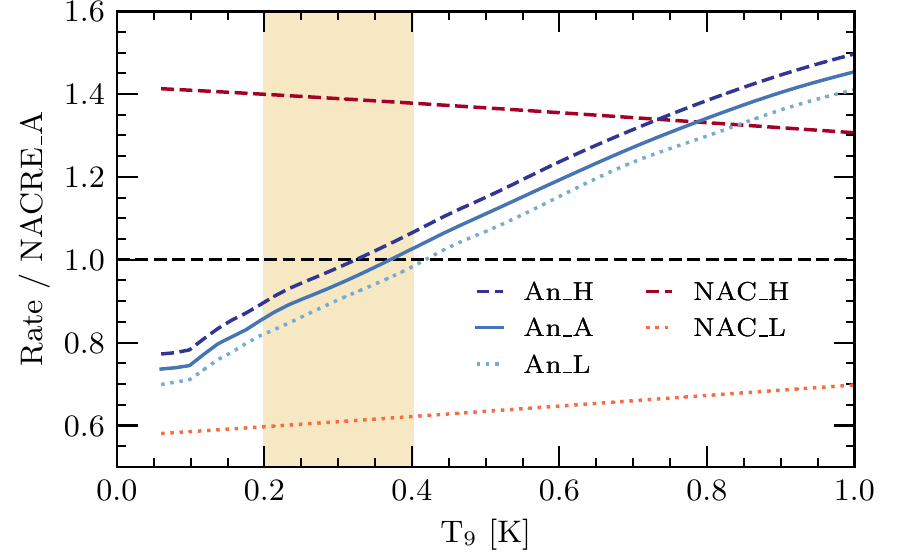}
    \caption{Ratios of each reaction rate considered when compared to the adopted NACRE rate for the $^{12}$C($\alpha$, $\gamma$)$^{16}$O reaction, as a function of temperature, where $T_9 = T / 10^9$. The beige shaded region defines the temperatures where helium burning occurs. During the core helium burning stage is also where the $^{12}$C($\alpha$, $\gamma$)$^{16}$O reaction is most prominent. The light-orange dotted and red dashed lines represent the NACRE\_L and NACRE\_H considerations, respectively. The solid blue line defines the adopted rate from \citet{an16} with the An\_L and An\_H rates being depicted as light-blue dotted and dark-blue dashed lines, respectively. }
    \label{fig: rate_ratios}
\end{figure}

Figure~\ref{fig: rate_ratios} shows a comparison between the adopted reaction rates from \citet{an16}, NACRE and all of their associated uncertainties. In this figure, we depict for each rate, the ratio between the rate and the value for NACRE\_A, as a function of temperature. For an analysis including other works, see Figure 4 of \citet{an16}. As can be seen from Figure~\ref{fig: rate_ratios}, for energies characteristic of stellar energies, the An\_A, An\_H and An\_L reaction rates are lower than for NACRE\_A for most temperatures within the blue shaded region, characteristic of core helium burning temperatures. We therefore expect to have a larger C/O ratio in the core after the central helium burning stage for the sequences which consider the rate from \citet{an16} when compared to those sequences which consider NACRE\_A. It can also be seen in Figure~\ref{fig: rate_ratios} that the range between NACRE\_H and NACRE\_L includes all the other prescriptions within the region of helium burning temperatures, which will lead to the largest differences in the C/O ratio after the core helium burning stage. At higher temperatures (greater than those considered to be helium burning temperatures) the reaction rate from \citet{an16} is larger than that from NACRE. These temperatures are not reached in the sequences computed within this work.

\section{Results and Discussions} \label{resdis}

In this section we describe in detail the effects that the uncertainties of the $^{12}$C($\alpha$, $\gamma$)$^{16}$O reaction rate have on the inner structure and evolution for low- and intermediate-mass single stars. As expected, during the pre-main sequence, main sequence (MS) and red-giant branch (RGB), we find no differences to the evolution since the $^{12}$C($\alpha$, $\gamma$)$^{16}$O reaction only becomes important, and increasingly more dominant, during the CHB as the central helium abundance decreases \citep{salaris05, spruit15, deboer19}. Thus, we report no difference between the different $^{12}$C($\alpha$, $\gamma$)$^{16}$O reaction rates at the time of, or shortly after, the helium-flash or a non-degenerate helium ignition. We only show the results from the CHB, AGB and WD stages where we expect some differences to occur due to the uncertainties and separate literature sources of the $^{12}$C($\alpha$,~$\gamma$)$^{16}$O reaction rate. We consider each evolutionary stage separately in chronological order.


\subsection{The Core Helium Burning Phase} \label{CHB}



\begin{figure}
    \centering
    \includegraphics[width=\linewidth]{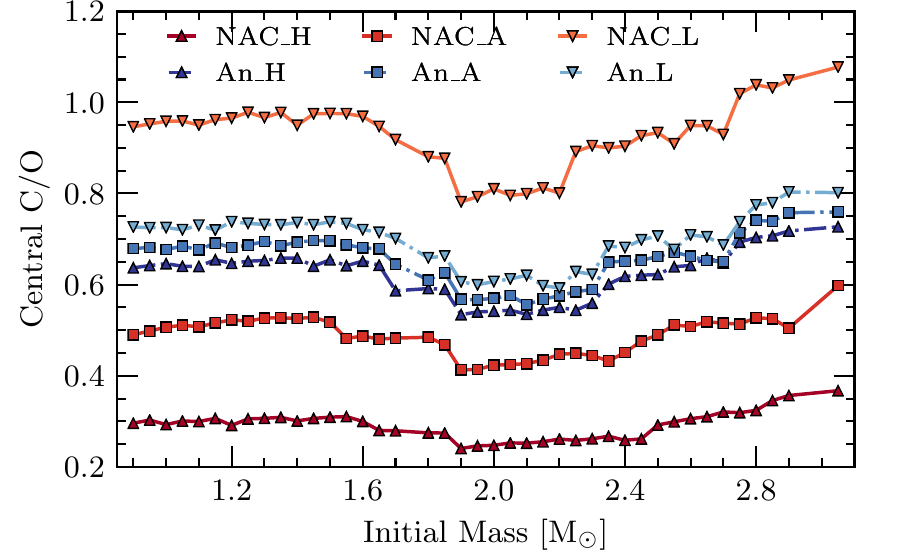}
    \caption{Central C/O ratio at the end of the CHB as a function of initial mass. The red points represent the reaction rates considered by NACRE and the blue points are those considered by \citet{an16}. Additionally, squares represent the respective adopted rates while darker-coloured triangles and lighter-coloured upside-down triangles represent the high and low limit uncertainties, respectively.}
    \label{fig: co_chb}
\end{figure}

Figure~\ref{fig: co_chb} shows the carbon--to--oxygen (C/O) ratio for each star at the end of CHB, as a function of initial mass. As expected due to the large uncertainties of the reaction rate from NACRE, the smallest and largest C/O ratios come from the NACRE\_H and NACRE\_L rates, respectively. Note that when all reaction rates from \citet{an16} are considered, the values for the C/O ratios are between the values corresponding to NACRE\_A and NACRE\_L. 


We find that the C/O ratio at the end of the CHB decreases for all considered reaction rates around an initial mass of $M_i = 1.90\, M_\odot$. This mass corresponds to the minimum mass for which helium burning starts in non-degenerate conditions, and will be referred to as the transition mass. The C/O ratio increases again for higher initial masses (between $2.20 \leq M_i/M_\odot \leq 2.45$). 
We find that the initial mass where the increase of the C/O ratio occurs is dependent on the considered $^{12}$C($\alpha$, $\gamma$)$^{16}$O reaction rate, such that higher reaction rates have a wider initial mass range for the decreased C/O ratio and lower reaction rates have a narrower initial mass range. For example, the NACRE\_H has the widest range ($1.90 \leq M_i/M_\odot \leq 2.45$) whereas the NACRE\_L has the narrowest range ($1.90 \leq M_i/M_\odot \leq 2.20$). Furthermore, we find no difference to the initial mass range between the adopted rate from \citet{an16} and the An\_H and An\_L rates. We also add that the decrease in the C/O ratio is more pronounced for less efficient reaction rates, see Figure~\ref{fig: co_chb}, for details.


\begin{figure*}
    \centering
    \includegraphics[width=\textwidth]{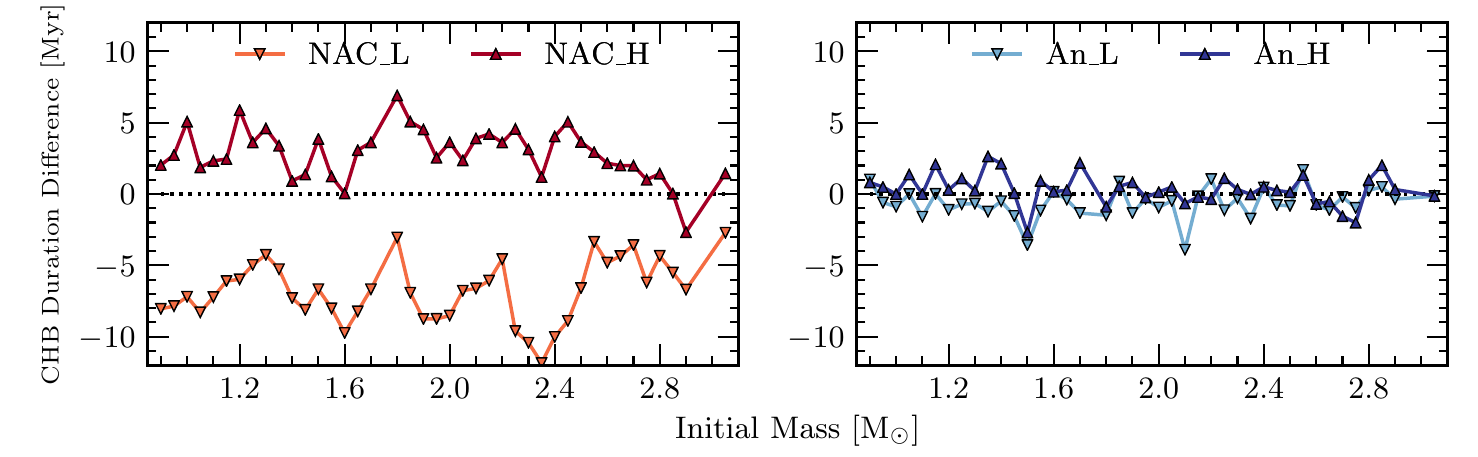}
    \caption{Differences to the duration of the CHB stage due to associated reaction rate uncertainties as a function of initial mass. The differences are calculated between each limit of the reaction rate due to their uncertainties and the adopted rate of each case. The left panel shows the differences of the uncertainties of the rate calculated by NACRE and the right panel shows the same from the rate of \citet{an16}. Darker-coloured triangles and lighter-coloured upside-down triangles represent the high and low limit uncertainties, respectively.}
    \label{fig: rel_chb_dur}
\end{figure*}

Figure~\ref{fig: rel_chb_dur} shows the time spent in the CHB as a function of initial mass for the High and Low reaction rate formulas for NACRE (left panel) and \citet{an16} (right panel). We consider the difference in the CHB age from the values obtained using the respective adopted reaction rate for each panel. Considering the NACRE rates (left panel of Figure~\ref{fig: rel_chb_dur}), we find that the CHB lifetime can be up to 12 Myr shorter (longer) from the adopted rate if we consider NACRE\_L (NACRE\_H) reaction rate, which is roughly a 7\% difference. On the other hand the differences between the \citet{an16} rates are much lower (right panel of Figure~\ref{fig: rel_chb_dur}), up to 4 Myr translating to a difference of 4\%. 
Such changes to the CHB lifetimes due to limits of the uncertainties on the $^{12}$C($\alpha$, $\gamma$)$^{16}$O reaction rate are not negligible, particularly for the rate taken from NACRE. \citet{constantino16} found that the difference in the the ratio of HB--to--AGB stars in a sample of 48 globular clusters could be explained by the differences in the CHB duration due to the uncertainties in the $^{12}$C($\alpha$, $\gamma$)$^{16}$O reaction rate.

\begin{figure}
    \centering
    \includegraphics[width=\linewidth]{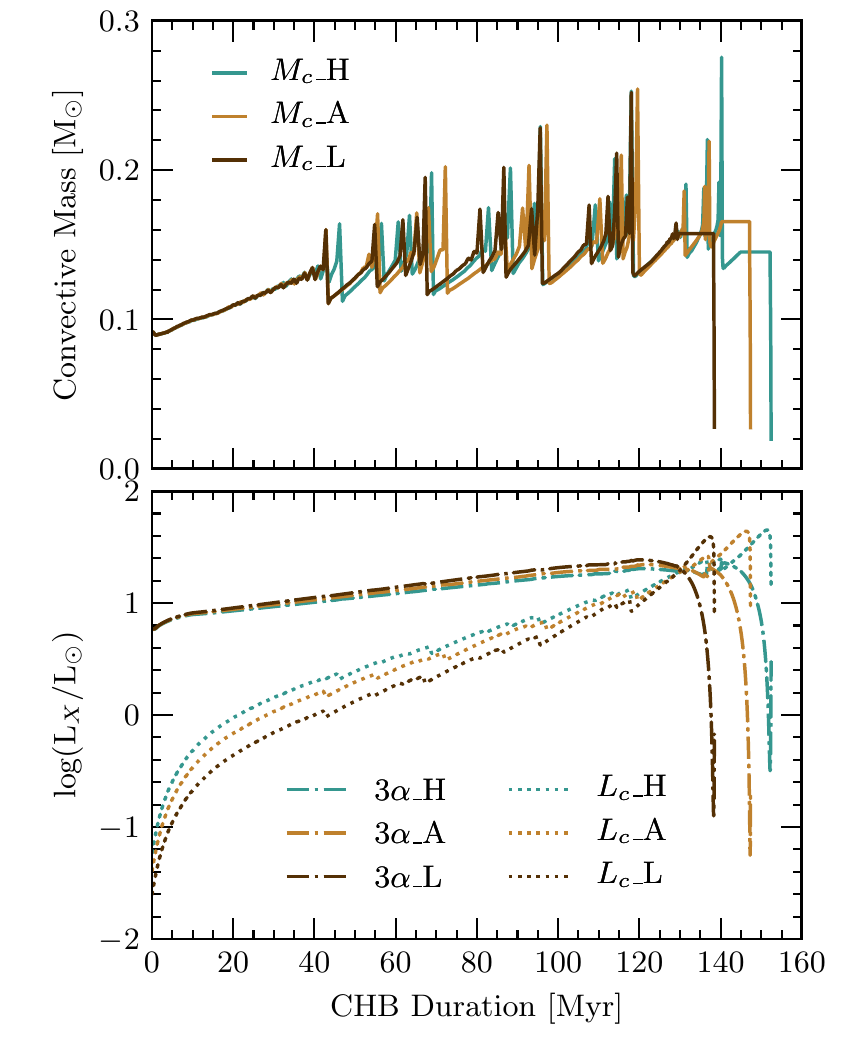}
    \caption{History of the convective mass (top panel), 3$\alpha$ luminosity and the luminosity of the $^{12}$C($\alpha$, $\gamma$)$^{16}$O reaction during the CHB (bottom panel). The history is given in terms of the CHB duration. This plot in particular considers all NACRE prescriptions for the $^{12}$C($\alpha$, $\gamma$)$^{16}$O reaction rate for an initial mass of $M_i = 2.45\, M_\odot$. Blue lines represent NACRE\_H, orange-brown depicts NACRE\_A and dark-brown shows NACRE\_L. Furthermore, the solid line represents the convective mass, dotted lines show the luminosity of the $^{12}$C($\alpha$, $\gamma$)$^{16}$O reaction and dot-dash lines portray the 3$\alpha$ luminosity.}
    \label{fig: ME}
\end{figure}

The top panel of Figure~\ref{fig: ME} shows the CHB history of the convective mass. The convective mass is defined as the mass-coordinate of the core convective boundary, such that convection occurs between this mass-coordinate and the centre. Additionally, the bottom panel of Figure~\ref{fig: ME} shows the luminosities of the 3$\alpha$ process and the $^{12}$C($\alpha$, $\gamma$)$^{16}$O reaction (the latter will be referred to as C$\alpha$ luminosity), for the NACRE reaction rates. As expected, the C$\alpha$ luminosity increases when the more efficient reaction rates are considered. Furthermore, the contribution from the 3$\alpha$ process decreases for higher reaction rates due to the helium reservoir being depleted faster by the more efficient $^{12}$C($\alpha$, $\gamma$)$^{16}$O reaction.

Mixing episodes due to the convective core during the CHB extends from the C/O core to the He-rich layers above, so we define the convective mass as the mass of the convective core. Figure~\ref{fig: ME} also shows that higher reaction rates produce more mixing episodes which are characterised by sudden increases of the convective mass. These enhanced convective episodes bring fresh helium from the helium region above the C/O core which not only increases the duration of the CHB but also increases the abundance of oxygen in the core \citep{ghasemi17, guo18}.

Convective mixing episodes induce a chemical discontinuity between the fully mixed core and the radiative layer, increasing the opacity beyond the convective boundary. In a class of CHB pulsating stars, sdB stars \citep[see][for an in depth discussion]{heber09}, g-modes propagate from the surface all the way until the boundary of the convective core \citep{ghasemi17}. Since we find significant differences to the size of the convective core and number of mixing episodes between the NACRE adopted reaction rate and its uncertainties for the $^{12}$C($\alpha$, $\gamma$)$^{16}$O reaction rate, the precision of astereoseismology for these objects is limited and must be considered in the calculations of the pulsation period spectrum. However for the adopted rate taken from \citet{an16}, the high and low limits (An\_H and An\_L, respectively) do not produce a significant change to the convective core mass and the total number of mixing episodes and would therefore produce a more precise study of the g-mode pulsations (see Figure~\ref{fig: ME_an} in Appendix~\ref{An_appendix}, for an example of the same case that considers the reaction rates from \citet{an16}). The implications for asteroseismology from the treatment to mixing during the CHB has been studied by \citet{constantino15} who found that changes to the composition and He-burning reaction rates do not significantly change the period spacing of pulsations for pulsators during the CHB stage. However, the period values could be more sensitive to the changes in the chemical profile.

\begin{figure}
    \centering
    \includegraphics[width=1.0\columnwidth]{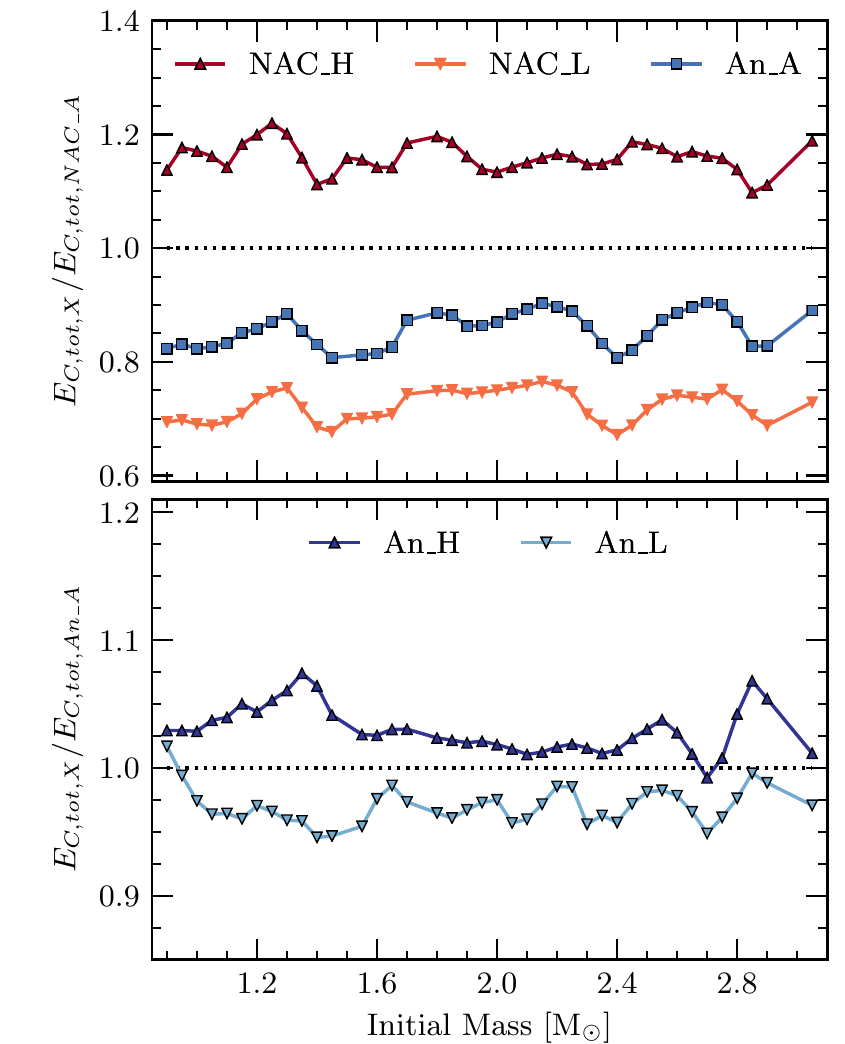}
    \caption{Ratios of the total energy produced by the $^{12}$C($\alpha$, $\gamma$)$^{16}$O reaction as a function of initial mass. Values are presented in the form of moving averages. The energy produced is calculated by integrating the C$\alpha$ luminosity shown in Figure~\ref{fig: ME} and is integrated with respect to time. The ratios in the top panel are in terms of the NACRE\_A rate and the ratios in the bottom panel are made in terms of An\_A. The red points represent the reaction rates considered by NACRE and the blue points are those considered by \citet{an16}. Additionally, squares represent the respective adopted rates while darker-coloured triangles and lighter-coloured upside-down triangles represent the high and low limit uncertainties, respectively.}
    \label{fig: Ec_chb}
\end{figure}

The total energy produced by the $^{12}$C($\alpha$, $\gamma$)$^{16}$O reaction during the CHB is presented in Figure~\ref{fig: Ec_chb}. The values shown in Figure~\ref{fig: Ec_chb} are moving averages. We compute the total energy by integrating the C$\alpha$ luminosity with respect to time for the CHB duration. Figure~\ref{fig: Ec_chb} shows the ratio between the different reaction rates and the NACRE\_A (top panel) and An\_A (bottom panel) reaction rates, as a function of initial mass. If we consider the reaction rates from \citet{an16}, the differences  are generally smaller than 10\%, the largest difference occurs for the sequence with an initial mass of $2.85\, M_\odot$ that considers An\_H. In most cases, the differences are no larger than 5\% (70.7\% of the sequences for An\_H and 82.9\% of the sequences for An\_L).

We find larger differences between the limits of $^{12}$C($\alpha$, $\gamma$)$^{16}$O NACRE rates when compared to the NACRE\_A formula, as shown in the top panel of Figure~\ref{fig: Ec_chb}. In this case we also compare the adopted reaction rate from \citet{an16}. If we consider how NACRE\_H differs from NACRE\_A, we find that the energy production for the majority of the sequences are greater than 10\% than that of the NACRE\_A case, with a few exceeding a difference of 20\%. For NACRE\_L, the carbon energy produced differs more than 30\% from the NACRE\_A rate. The extra energy produced from the high rates when compared to the adopted rates increases the temperature gradient further allowing convection to continue, causing the extra mixing episodes shown in Figure~\ref{fig: ME} \citep{kipp90, prialnik09}. 

Considering the adopted rate from \citet{an16}, the absolute value of the differences in carbon energy produced due to An\_H and An\_L appears to be independent of either selection. This is not the case for the NACRE rates. A limiting factor for the amount of energy produced is the abundance of available helium. This is more of a limit for the NACRE\_H case due to lack of available helium inhibiting further reactions to occur. The NACRE\_L will always produce less carbon energy and so is not limited by the helium abundance or lack thereof. The smaller uncertainties of the rates taken from \citet{an16} are not large enough to produce such an effect.

The CHB stage is where the $^{12}$C($\alpha$, $\gamma$)$^{16}$O reaction is the most active. In particular, we find that the largest differences due to the considered $^{12}$C($\alpha$, $\gamma$)$^{16}$O reaction rate appear in the final C/O ratio, CHB duration, energy generation rate and the number of experienced mixing episodes. The primary reason that we find such changes to these properties is due to the changes in energy generation that affects the convection efficiency in this phase. 
Furthermore, we find that the differences between the An\_H and An\_L rates from the An\_A rate are generally insignificant, unlike those of the NACRE uncertainties which are intrinsically larger. A final point to add is that, in future works, the use of overshooting parameters specifically designed for the CHB would be interesting. Works such as \citet{spruit15} claim to keep the convective boundaries stable inhibiting the need for manual breathing pulse suppression, as performed in this work, whilst keeping "stable" convection active throughout the evolution \citep{spruit15, constantino17}.

\subsection{The Asymptotic Giant Branch Phase} \label{AGB}

During the AGB the energy production is given by two shell sources, the hydrogen-shell at the base of the hydrogen-rich envelope and the He-shell on top of the C/O core. Hydrogen burning occurs through the CNO cycle, while He-burning is through the 3$\alpha$ process. Towards the end of the AGB, the He-burning shell will become thin enough to trigger unstable burning, and the thermal pulses (TPs) begin \citep[e.g.][]{kipp90, iben91}. During the interpulse period between the TPs, the outer convection zone may be deep enough to bring the products of He-shell burning to the surface, this is known as the third dredge-up (TDU) \citep{wallerstein97, busso99, herwig05, karakas14}.

Well known consequences of TDUs are a reduction of the helium core mass and changes to the surface composition, leading to the formation of C-stars \citep[][]{frost96, busso99, karakas02, weiss09, romero15, marigo20}. The extent of the reduction of the helium core mass from TDU episodes is parameterised by the dredge-up efficiency parameter, $\lambda_d$\footnote{The dredge-up efficiency parameter is defined as the fraction of helium core mass lost during the TDU episode over the helium core mass growth since the last TDU} \citep[see][for details]{karakas02, marigo13}. The $^{12}$C($\alpha$,~$\gamma$)$^{16}$O reaction during this stage is essentially inactive. There may be some fusion reactions between $^{12}$C and alpha particles at the edge of the C/O core but they are, however, insignificant. Thus, any difference between the sequences during the AGB is due to the effect that the $^{12}$C($\alpha$, $\gamma$)$^{16}$O reaction rate has during the CHB.

\begin{figure}
    \centering
    \includegraphics[width=1.0\columnwidth]{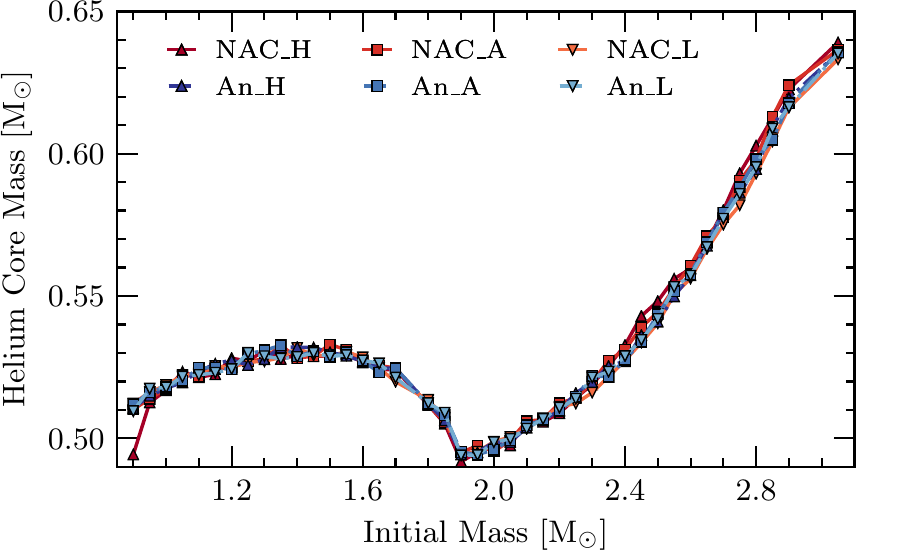}
    \caption{Helium core mass at the start of the first TP as a function of initial mass. All of the considered reaction rates and their uncertainties are shown within this figure. We find a minimum to the helium core mass for the same initial mass which corresponds to the transition mass where core helium burning begins on a non-degenerate core rather an electron degenerate core ($M_i = 1.90\, M_\odot$). Within the uncertainties, we find differences up to 0.01 M$_\odot$ for masses larger than $M_i = 1.90\, M_\odot$. The red points represent the reaction rates considered by NACRE and the blue points are those from \citet{an16}. Additionally, squares represent the respective adopted rates while darker-coloured triangles and lighter-coloured upside-down triangles represent the high and low limit uncertainties, respectively.}
    \label{fig: Mc-1TP}
\end{figure}

Figure~\ref{fig: Mc-1TP} shows the helium core mass at the first TP of each sequence as a function of initial mass. A minimum value occurs for an initial mass $M_i = 1.90\, M_\odot$, which is the transition point as described in Section~\ref{CHB}. The same result was found in the work of \citet{kalirai14}, whose initial models come from those produced in \citet{bressan12}. However, their transition point occurs for $M_i = 2.00\, M_\odot$ due the larger initial metallicity affecting the mass for which core helium burning ignites in degenerate conditions \citep{bertelli86, romero15}. We find that there is no significant difference to the helium core mass at the first TP as a result of different $^{12}$C($\alpha$, $\gamma$)$^{16}$O reaction rates for masses lower than the transition point. Above this mass, the maximum difference between the NACRE rates is $\sim 0.01M_\odot$, with NACRE\_L producing lower helium core masses and NACRE\_H producing larger helium core masses. This is due to the difference in energy outputs between the adopted rate, NACRE\_A, and the NACRE\_H/NACRE\_L rates. Higher reaction rates during the CHB increase the temperature throughout the star which favours the CNO-cycle \citep[][]{boeltzig16}, allowing the helium core mass to develop further than sequences which consider lower reaction rates. There are no significant differences in the helium core mass at the first TP between the adopted rate from \citet{an16} and An\_H/An\_L for any of the considered initial masses.

\begin{figure}
    \centering
    \includegraphics[width=1.0\columnwidth]{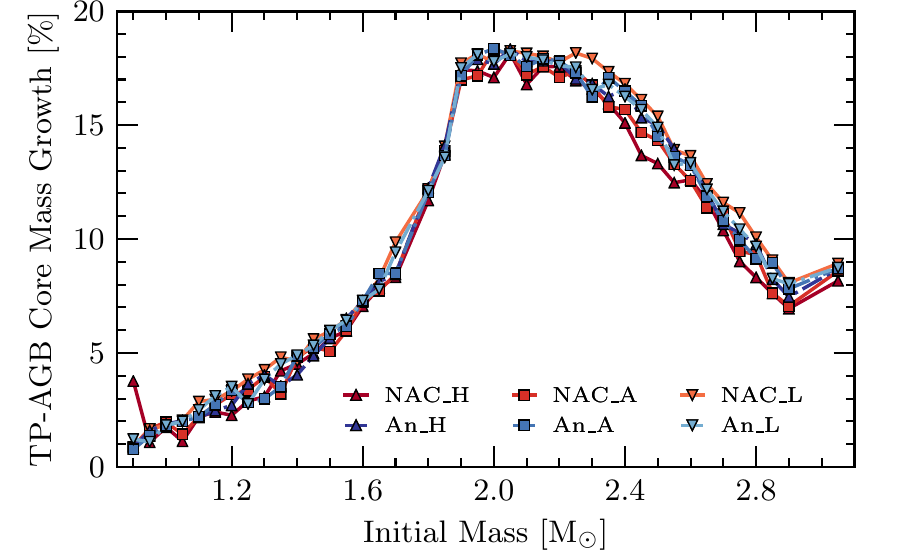}
    \caption{Percentage growth of the helium core mass during the AGB as a function of initial mass. Growth is calculated as the difference between the final mass of the core and the helium core mass described in Figure~\ref{fig: Mc-1TP}. We find that the largest growth occurs for initial masses $\approx 2.00\, M_\odot$, peaking at 19\%. Above initial masses of $M_i = 2.90\, M_\odot$, it appears that the growth begins to plateau around 8-9\%. The red points represent the reaction rates considered by NACRE and the blue points are those from \citet{an16}. Additionally, squares represent the respective adopted rates while darker-coloured triangles and lighter-coloured upside-down triangles represent the high and low limit uncertainties, respectively.}
    \label{fig: core_growth}
\end{figure}

Figure~\ref{fig: core_growth} shows the growth of the helium core mass during the TP-AGB as a function of initial mass for each considered reaction rate. We find that the dramatic increase of core growth (for helium core mass growth $\geq 10\%$ \citep{kalirai14}) occurs in the range $1.70 \leq M_i/M_\odot \leq 2.60$, with a maximum increase of 19\% occurring at $M_i \approx 2.00\, M_\odot$. This result is in agreement with that of \citet{bird11} and is similar to that of \citet{kalirai14}, who find a helium core growth up to 30\%. This discrepancy between their work and ours is due to not only a different initial metallicity, but also their consideration of a less efficient mass-loss scheme for stages previous to the AGB (Reimers law with $\eta_R = 0.2$ \citep{bressan12}). Thus, the models used by \citet{kalirai14} have a larger mass of hydrogen fuel to produce a larger final mass (see Table~\ref{tab:my-table} for our values of this variable and \citet{bird11} for an in-depth discussion of the hydrogen fuel variable). Furthermore, possible differences to the energy produced in the H-rich envelope during the TP-AGB may affect the rate of the helium core growth \citep[see][for details]{forestini97, marigo13, kalirai14}.

Considering only the difference in helium core mass growth for NACRE\_A rate and it's NACRE\_H/NACRE\_L limits, we find that NACRE\_L has a larger core growth and NACRE\_H has smaller core growth. The increased core growth during the AGB for the NACRE\_L sequences is due to the smaller helium core mass at the first TP (see Figure~\ref{fig: Mc-1TP}) and as such more fuel to keep He-shell burning sustained, particularly for initial masses above the transition point where the core growth differences are greater (see Table~\ref{tab:my-table}). Additionally, during the TP-AGB, we find differences in the energy generation from the CNO cycle between the NACRE\_H/NACRE\_L limits in comparison with the NACRE\_A. The energy generation can be up to 25\% lower (higher) when the NACRE\_H (NACRE\_L) reaction rate is considered.


\begin{figure}
    \centering
    \includegraphics[width=1.0\columnwidth]{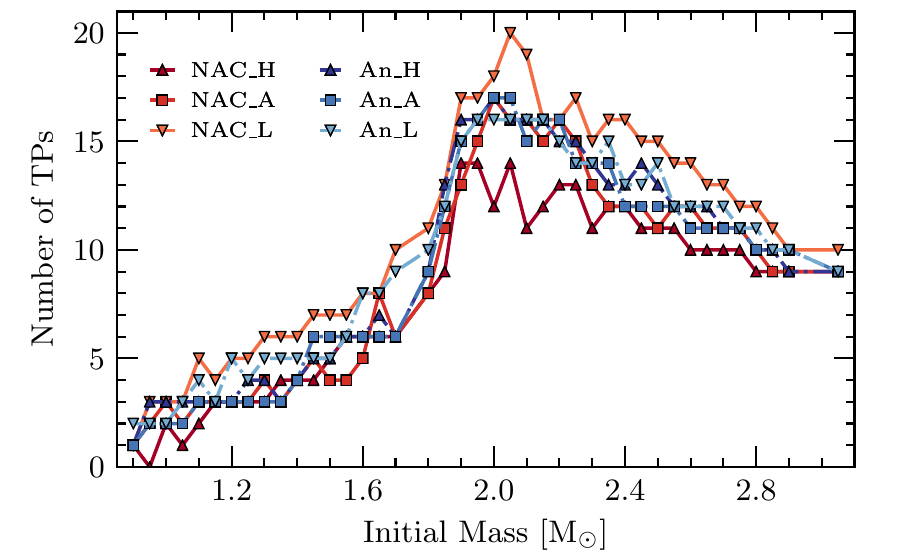}
    \caption{Number of TPs experienced as a function of initial mass. Each reaction rate consideration and their uncertainties are shown. We find that the number of TPs peaks at initial masses $\approx 2.00\, M_\odot$, in-line with the largest core growth, as in Figure~\ref{fig: core_growth}. We also show that lower reaction rates for the $^{12}$C($\alpha$, $\gamma$)$^{16}$O reaction produce more TPs. The red points represent the reaction rates considered by NACRE and the blue points are those from \citet{an16}. Additionally, squares represent the respective adopted rates while darker-coloured triangles and lighter-coloured upside-down triangles represent the high and low limit uncertainties, respectively.}
    \label{fig: no_tps}
\end{figure}

Figure~\ref{fig: no_tps} shows the number of thermal pulses as a function of initial mass for each considered reaction rate. Moreover, it shows that lower reaction rates experience more TPs than higher reaction rates. This is related to the larger amount of available hydrogen to aid the outward growth of the helium core through a greater number of unstable He-shell burning episodes - TPs. We do not find any M-star to C-star transitions \citep[see][for example]{marigo20} as convective overshooting about the boundary between the helium core and the He--exhausted core was disregarded during the TP-AGB, inhibiting the TDU \citep{herwig2000, romero15}. However, overshooting still occurred at the boundary of the H--rich core. We define the "He--exhausted core" as the region from the centre until the local abundance of helium is greater than $10^{-6}$.

Thermal pulses are strongly dependent on the mass--loss rate, helium core mass and initial metallicity \citep{karakas02, cristallo09, weiss09, renedo10, romero15, DG17}. We find that the number of thermal pulses in our computations is lower than that from the works of  \citet{weiss09, renedo10} and \citet{romero15} for a given initial mass, a similar treatment of convection and a similar helium core mass at the beginning of the TP-AGB phase. Difference in the number of TPs could be related to the different mass--loss schemes during the RGB stage. In this work we consider the mass--loss prescription from \citet{bloecker95} while the works of \citet{weiss09, renedo10} and \citet{romero15} consider a mass--loss scheme that produces a "super wind" stage towards the last TPs, making it more efficient in these last TPs but less so in the early TP-AGB \citep[see][for details]{VW93, vl05}. However, the trend in the number of experienced TPs as a function of initial mass obtained in our work agrees with other works \citep[see][]{weiss09, renedo10, romero15}.

To assess the effect of the TDU during the TP-AGB, we computed additional sequences, allowing convective overshooting to occur at all fully-- or semi--convective boundaries, with $f = 0.016$ (see Appendix~\ref{model_appendix}, for details on it's effect). For sequences that consider the NACRE\_A prescription, TDU episodes occur for initial masses larger than $M_i \geq 2.40\;M_\odot$, with the dredge-up efficiency parameter ($\lambda_d$) showing values of $\lambda_{d} = 0.033 - 0.124$ that increases with increasing initial mass. The abundance of carbon and oxygen at the surface does increase during each TDU in these additional models, but the C/O is still lower than 1 meaning that our models show an oxygen dominated surface. A higher value of the overshooting parameter may be necessary to produce C--stars \citep[see][for examples of C-star transitions]{herwig97, karakas02, weiss09, romero15, marigo20}. 
For sequences where convective overshooting was considered across all boundaries during the AGB we find a decrease in the final helium core mass up to 0.63\%. This value is much lower than the 15\% decrease found by \citet{karakas02, romero15}.

The sequences that have initial masses $M_i < 2.40\;M_\odot$ do not show any third dredge--up episodes, as such we do not expect any difference to the growth of the helium core or the final mass. For those sequences with initial masses $M_i \geq 2.40\;M_\odot$, a more detailed study of the convective boundaries during the TP-AGB is required for more thorough analysis of why we find such weak dredge--up efficiency parameters.

In the case of NACRE\_H and NACRE\_L, we find that TDU episodes occur for the same initial mass range as that of the NACRE\_A sequences ($2.40 \leq M_i/M_\odot \leq 3.05$). Additionally, the dredge-up efficiency parameters are also similar to those of the NACRE\_A sequences, with $\lambda_d = 0.040 - 0.123$. From the results gathered in this work, we find that the uncertainties of current $^{12}$C($\alpha$, $\gamma$)$^{16}$O reaction rates are not significant in modelling the TDU.

The $^{12}$C($\alpha$, $\gamma$)$^{16}$O reaction during the AGB is negligible during the TP-AGB. Instead, the main energy source occurs through the 3$\alpha$ reaction series and the CNO-cycle within the H-rich envelope \citep{herwig05, karakas14}. Thus, we do not find any significant change to the peak TP luminosity nor the depth of each TDU, since the changes in core mass at the beginning of the TP-AGB are negligible as a result of the uncertainties of the $^{12}$C($\alpha$, $\gamma$)$^{16}$O reaction rate, as shown in Figure~\ref{fig: Mc-1TP} \citep[see][for details]{wallerstein97, wagenhuber98, busso99, herwig05, karakas14}. 
However, the uncertainties of the overshooting efficiency raises a greater uncertainty in the surface composition during the AGB, as such we leave a detailed discussion for a future work that considers the overshooting efficiency in more detail \citep{abia02, herwig05, cristallo09, ventura09, karakas14}.

\begin{table*} 
\begin{tabular}{|c|c|c|c|c|c|c|c|c|}
\hline
\multirow{2}{*}{$M_i$/$M_\odot$} & \multicolumn{4}{c|}{$\Delta M_{\text{growth}}/M_\odot$} & \multicolumn{4}{c|}{$M_{\text{fuel}}/M_\odot$}        \\ \cline{2-9} 
                       & NACRE\_H  & NACRE\_A  & NACRE\_L & An\_A & NACRE\_H & NACRE\_A & NACRE\_L & An\_A \\ \hline
1.00                   & 0.009     & 0.010     & 0.009    & 0.009 & 0.007    & 0.008    & 0.007    & 0.008 \\ \hline
1.50                   & 0.030     & 0.027     & 0.031    & 0.031 & 0.024    & 0.022    & 0.026    & 0.025 \\ \hline
1.60                   & 0.037     & 0.038     & 0.039    & 0.038 & 0.030    & 0.031    & 0.031    & 0.031 \\ \hline
2.00                   & 0.085     & 0.091     & 0.089    & 0.091 & 0.069    & 0.073    & 0.072    & 0.073 \\ \hline
2.90                   & 0.043     & 0.044     & 0.050    & 0.048 & 0.035    & 0.035    & 0.040    & 0.039 \\ \hline
\end{tabular}
\captionsetup{justification=centering}
\caption{Values showing the TP-AGB helium core mass growth and fuel mass. We report the values from the following reaction rate considerations: NACRE\_H, NACRE\_A, NACRE\_L and An\_A. We do not report the values from the uncertainties of the rate taken from \citet{an16} since they are negligible when compared to their adopted rate.}
\label{tab:my-table}
\end{table*}

\subsection{The White Dwarf Final Cooling Track} \label{WD}



\begin{figure}
    \centering
    \includegraphics[width=1.0\columnwidth]{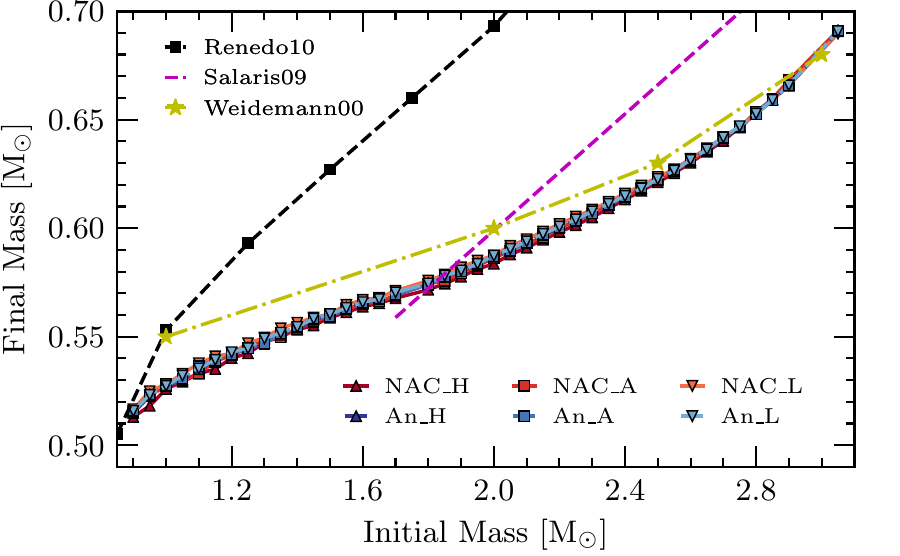}
    \caption{Initial-to-final mass relation of all sequences calculated as part of this work. Also shown are other IFMRs from the works of \citet{weidemann00, salaris09, renedo10} (yellow stars, purple dashed line and black squares, respectively) for a comparison of their trends. The red points represent the reaction rates considered by NACRE, and the blue points are those from \citet{an16}. Additionally, squares represent the respective adopted rates while darker-coloured triangles and lighter-coloured upside-down triangles represent the high and low limit uncertainties, respectively. We find that the slope of the IFMR has a strong dependency on the considered mass-loss scheme considered during the AGB, with the scheme from \citet{VW93} producing a steeper gradient and that from \citet{bloecker95} showing a shallower gradient.}
    \label{fig: ifmr}
\end{figure}

Figure~\ref{fig: ifmr} shows the initial-to-final mass relation (IFMR) for all sequences produced in this work. We find that there is no significant difference in the final mass of any given initial mass due to the $^{12}$C($\alpha$, $\gamma$)$^{16}$O reaction rate. Considering the largest difference in the reaction rates, between NACRE\_H and NACRE\_L, the largest difference in the final mass for a given initial mass is less than $0.01\, M_\odot$ ($<2\%$).

In the interest of the pursuit for a global IFMR, we compare our IFMR to those of other works of a similar metallicity. We consider the IFMRs from the works of \citet{weidemann00, salaris09} and \citet{renedo10}. We find a similar trend with the work of \citet{weidemann00}, both of which consider the same mass-loss scheme from \citet{bloecker95} for the AGB phase. The IFMRs from the works of \citet{salaris09} and \citet{renedo10} consider the mass--loss scheme from \citet{VW93} for the AGB and show a much steeper gradient in their IFMRs. However, the core masses between this work and the works of \citet{weidemann00, salaris09} and \citet{renedo10} are similar at the first TP. Thus, it is reasonable to assume that the difference is due to their considered mass-loss scheme for the IFMR determination. 

By considering the third-order polynomial nature of the IFMR computed in this work, we fit a function to the NACRE\_A final masses to produce a general relation from the results of this work. This allows for a comparison to other IFMRs as well as other masses to be easily estimated, if desired. The following IFMR reproduces the IFMR of NACRE\_A well, such that the R-square value is $R^2 = 0.9995$:

\begin{equation} \label{eq: my_ifmr}
    M_f = 0.02047M_i^3 - 0.1051M_i^2 + 0.2323M_i + 0.3783M_\odot
\end{equation}

\noindent where $M_f$ is the final mass and $M_i$ is the initial mass. 
The non-linear relationship described by Equation~\ref{eq: my_ifmr} is caused by the mass-loss rate adopted on the AGB. The \citet{bloecker95} scheme in particular has a large dependency on luminosity. It would be interesting to see how our IFMR holds for observational data as well as it's dependency on metallicity - an important dependence as discussed in \citet{romero15}.

\begin{figure*}
    \centering
    \includegraphics[width=1.0\linewidth]{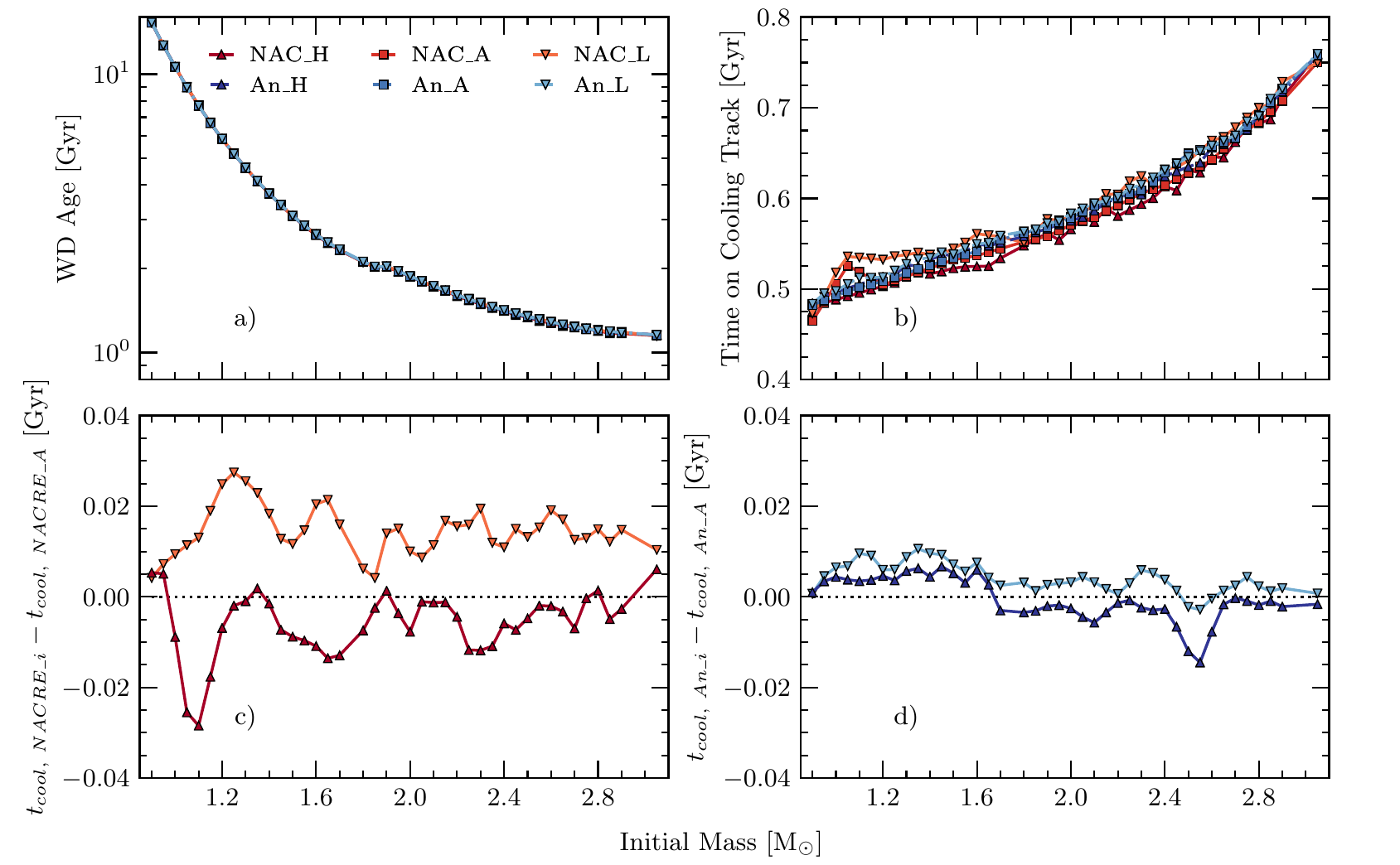}
    \caption{Panel a) shows the final age (log scale) of the star on the final cooling track with an effective temperature $T_{\text{eff}} = 10\, 000$K. Panel b) shows the time spent of the cooling track, defined as the time taken for a WD on the final cooling track to cool from its maximum effective temperature to an effective temperature of $T_{\text{eff}} = 10\, 000$K. Panel c) and d) show the moving average for the difference of cooling times between the High/Low limits and the Adopted rate for the NACRE and \citet{an16} $^{12}$C($\alpha$, $\gamma$)$^{16}$O reaction rate, respectively. All panels are represented as functions of initial mass. The NACRE reaction rates are shown as different shades of red and those from \citet{an16} are depicted by shades of blue. Furthermore, squares represent the respective adopted rates while darker-coloured triangles and lighter-coloured upside-down triangles represent the high and low limit uncertainties, respectively. In general, we find that the uncertainties of the $^{12}$C($\alpha$, $\gamma$)$^{16}$O reaction rate have an negligible effect on the final ages of the stars at this point, whereas the cooling time can differ up to 8\%.}
    \label{fig: age}
\end{figure*}

In Figure~\ref{fig: age} we show, in panel a), the final ages of a WD that has cooled to an effective temperature of $T_{\textrm{eff}} = 10\,000$K (log scale) as a function of initial mass for all the sequences computed in this work. The differences in the final ages due to the High/Low limits of each considered $^{12}$C($\alpha$, $\gamma$)$^{16}$O reaction rate are in general negligible, with variations of the order $\sim 0.01$ Gyr for both the NACRE and \citet{an16} $^{12}$C($\alpha$, $\gamma$)$^{16}$O reaction rate. The variations in the reported final ages due to the uncertainties of the $^{12}$C($\alpha$, $\gamma$)$^{16}$O reaction rate are a  magnitude lower than the populations studied in the works of \citet{hansen13, forbes15, campos16}. As such, the impact that the $^{12}$C($\alpha$, $\gamma$)$^{16}$O reaction rate has on final ages of WD models is currently negligible as compared to the greater uncertainty of ageing stellar populations. 

Panels c) and d) of Figure~\ref{fig: age} show the moving average for the time spent on the cooling track for the NACRE and \citet{an16} $^{12}$C($\alpha$, $\gamma$)$^{16}$O reaction rates, respectively. We define this quantity as the time taken for a star on the final cooling track to cool from it's maximum effective temperature until an effective temperature of $T_{\text{eff}} = 10\, 000$K. During the final cooling track, the differences in the duration due to the reaction rates between the Adopted and High/Low limits generally differ up to $0.030$ Gyr for those of NACRE and up to $0.015$ Gyr for \citet{an16}. The general trend is in agreement with past discussions of the effect of the $^{12}$C($\alpha$, $\gamma$)$^{16}$O reaction rate and cooling time during this stage of evolution, such that more oxygen-rich cores will produce a lower cooling time. This is due to the gravitational energy release during stratification occurring at earlier times for more oxygen-rich cores. As a consequence, the WD is left with a lower thermal content to feed the surface luminosity at later times. The larger the luminosity at which the stratification occurs, the shorter the resulting cooling times will be \citep{dantona90, prada02, salaris10}. Furthermore, for the High/Low limits of the NACRE rate, we find that NACRE\_L produces a greater absolute difference than that of NACRE\_H. This is due to the availability of helium during the CHB as discussed in Section~\ref{CHB}. 

\begin{figure}
    \centering
    \includegraphics[width=1.0\columnwidth]{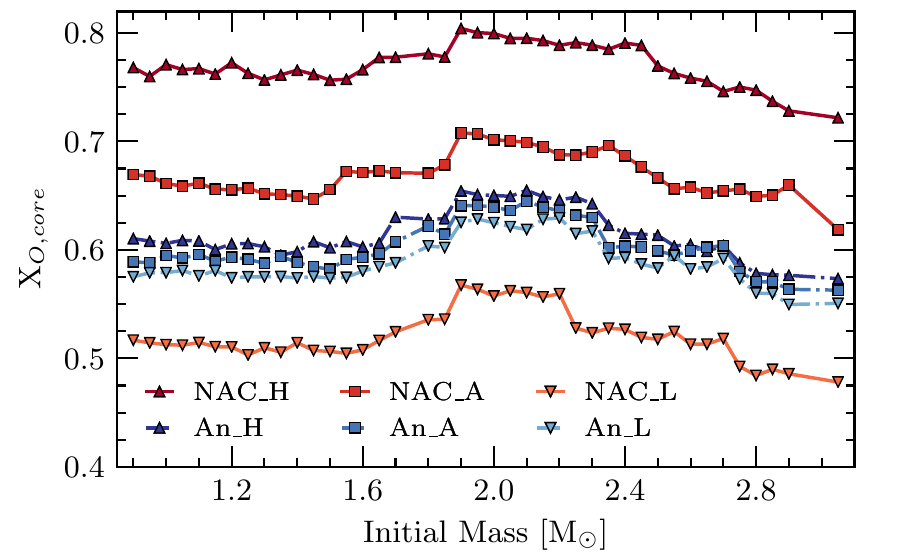}
    \caption{Central oxygen mass fraction for the final WD as a function of initial mass. We show each calculated sequence. The trends for each considered reaction rate are similar to those found in Figure~\ref{fig: co_chb}. There has been a slight increase in the central oxygen abundance since the CHB due to diffusion processes in the star. Additionally, squares represent the respective adopted rates while darker-coloured triangles and lighter-coloured upside-down triangles represent the high and low limit uncertainties, respectively.}
    \label{fig: o_dav}
\end{figure}

After the settling and diffusion processes described in Section~\ref{CT}, the final oxygen abundances within the core of the sequences are presented in Figure~\ref{fig: o_dav}, as a function of initial mass. We find similar trends to the oxygen mass fraction in this stage to those found at the end of the CHB. Although there are slight increases to the oxygen mass fraction due to the aforementioned diffusion processes \citep{unglaub00}. Additionally, diffusion affects the C/O ratio throughout the star up to the surface and not just in the core \citep[see][for details]{herwig2000, straniero03}.

The onset of crystallisation begins when the core cools to a certain temperature, $T_c$ \citep{segretain94, horowitz10}. This temperature is dependent on the internal composition of the star. Through observations of the globular cluster NGC 6397, \citet{winget09} report that the crystallisation of the WD core is similar to that of a pure carbon core. According to the phase diagram produced in \citet{horowitz10} and their limits for the maximum crystallisation temperature, this would require a limit to the oxygen mass fraction of $X_{\text{O}} \leq 0.64$. This requires that the maximum S-factor at 300 keV has an upper limit of $S(300\, \text{keV}) \leq 170\, \text{keV b}$. Considering the relationship between oxygen mass fraction and initial mass presented in Figure~\ref{fig: o_dav}, we find that NACRE\_H and NACRE\_A produce central oxygen abundances that are too large for a crystallisation process similar to that found by \citet{horowitz10}. Meanwhile, the rates \citet{an16} agree not only with the oxygen mass fraction limit presented by \citet{horowitz10}, but also their derived S-factor for an energy of 300 keV. Thus, we find that sequences dedicated to studying crystallisation using the method presented by \citet{horowitz10} should consider a lower reaction rate than that from NACRE for the $^{12}$C($\alpha$, $\gamma$)$^{16}$O reaction to keep their analysis consistent with the input physics that they use.


\begin{figure}
    \centering
    \includegraphics[width=1.0\columnwidth]{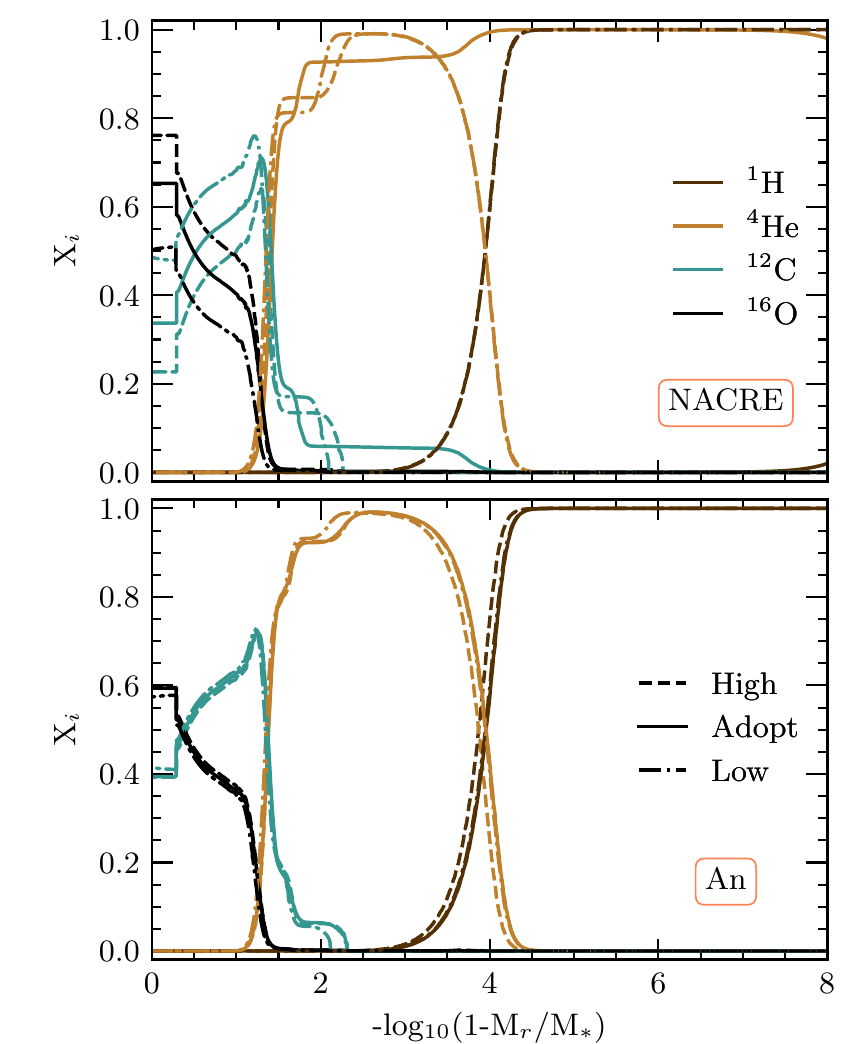}
    \caption{In both panels we show the abundance profiles of sequences considering an initial mass of $M_i = 1.30\, M_\odot$. The top panel represents the adopted rate and it's uncertainties for the NACRE rate, and the same for the \citet{an16} rates in the bottom panel. The line-styles for each rate are shown in the legend in the bottom panel and the colours for each element is shown in the legend in the top panel. Colour version is available online.}
    \label{fig: profile}
\end{figure}

Figure~\ref{fig: profile} shows the abundance profiles of white dwarf models with a stellar mass of $M_* = 0.548M_\odot$, $T_{\textrm{eff}} = 20\,000$K and an initial mass of $M_i = 1.30 \, M_\odot$. Sequences that consider a reaction rate from NACRE are shown in the top panel and those from \citet{an16} are represented in the bottom panel. All sequences finish with similar structure to those shown in Figure~\ref{fig: profile}. The profiles depict a DA white dwarf configuration, with a hydrogen-rich envelope, a helium buffer and a C/O core. 
Where the abundance of carbon reaches it's maximum, we hereafter refer to this as the carbon peak. 

We show that the interior of the star has a consistent trend where the carbon peak is higher for lower reaction rates - an outcome of a less efficient reaction rate which leaves behind a larger abundance of carbon. Furthermore, the position of the carbon peak changes with the reaction rates, moving away from the centre as the reaction rate increases.
We find in general that differences between An\_A and the An\_H/An\_L reaction rates do not affect this region drastically (bottom panel), unlike that of the NACRE $^{12}$C($\alpha$, $\gamma$)$^{16}$O reaction rate considerations (top panel).

The abundance profile and composition gradients in these central regions that lie within the range of $1 < -\textrm{log}_{10}(1 - M_r/M_*) < 2$ affect the peaks in the Brunt-V\"{a}is\"{a}l\"{a} frequency, which disturbs the period spectrum structure \citep[see][for more details]{corsico06, romero2012}. This is an outcome of the pulsation modes that are trapped in this region through the mode-trapping mechanism. We confirm that uncertainties of the $^{12}$C($\alpha$, $\gamma$)$^{16}$O reaction rate may affect the pulsation period spectrum. Another region where the Brunt-V\"{a}is\"{a}l\"{a} frequency is affected is in the He/H transition region. In particular, the position of the He/H transition will impact the period spectrum \citep{romero12, romero13}.

\section{Conclusions} \label{conc}

In this work we analyse the impact that the limits of the $^{12}$C($\alpha$, $\gamma$)$^{16}$O reaction rate has on the inner structure and evolutionary properties of low- and intermediate-mass stars. We consider the $^{12}$C($\alpha$, $\gamma$)$^{16}$O reaction rates from NACRE \citep{angulo99} and \citet{an16}. We have computed stellar sequences from the ZAMS until the remnant white dwarf reaches a luminosity of $\text{log}(L/L_\odot) = -3$. We applied similar starting parameters for different ensembles of reaction rates where we consider the adopted rate along with the upper and lower limits within the uncertainties of each source. We summarise our main results below.

\begin{enumerate}
  
  \item The C/O ratio of the core in the final model of each sequence is affected by the $^{12}$C($\alpha$, $\gamma$)$^{16}$O reaction rate as expected, with lower C/O ratios for larger reaction rates. We find that the decreased C/O ratio for initial masses greater than the transition mass increase again at higher masses. The mass at which this increase occurs is dependent on the considered $^{12}$C($\alpha$, $\gamma$)$^{16}$O reaction rate, such that it occurs for higher masses if higher reaction rates are considered. This is due to an increased number of mixing episodes, a cause of larger energy outputs increasing convective efficiency which brings fresh helium to the core during the CHB. Note that significant differences between the adopted rate and high/low limits occur only for those rates taken from NACRE which has a much larger uncertainty than those from \citet{an16}.
  
  \item CHB lifetime is dependent on the considered reaction rate, a higher reaction rate produces a greater lifetime. We deem this to be a consequence in the number of mixing episodes extending the core helium burning lifetime, although further research would be beneficial to confirm this. Between the adopted rate and high/low limits, we find a difference up to 12 Myr for the NACRE rates and up to 4 Myr for those from \citet{an16}. 
  
  \item The helium core mass at the beginning of the first TP is independent of the considered $^{12}$C($\alpha$, $\gamma$)$^{16}$O reaction rate up to and including the transition mass. Above this mass, we find a maximum difference of $\approx 0.01M_\odot$ between NACRE\_H and NACRE\_L, with lower reaction rates producing a lower helium core mass. Additionally, our minimum helium core mass at this point occurs at our transition mass.
  
  \item Growth of the helium core mass between the first TP and the final mass reaches a maximum of 19\%, with growths greater than 10\% occurring in the mass range $1.70 \leq M_i/M_\odot \leq 2.60$ which is in agreement with \citet{bird11} and \citet{kalirai14}. The largest growths occur for the lower reaction rates due to more available hydrogen which remained after the CHB. There are no significant differences between the rates taken from \citet{an16} due to the limits being smaller in relation to their adopted rate than those from NACRE.
  
  \item The number of TPs during the TP-AGB is dependent on the considered $^{12}$C($\alpha$, $\gamma$)$^{16}$O reaction rate. We find that lower reaction rates increase the number of TPs due to a larger hydrogen fuel aiding the outward growth of the helium core mass by fuelling the unstable He-shell with a greater supply of fresh helium.
  
  \item TDU episodes occur for sequences in the initial mass range of $2.40 \leq M_i/M_\odot \leq 3.05$ with dredge-up efficiency parameters $\lambda_d = 0.033 - 0.124$. This mass range is independent of the considered $^{12}$C($\alpha$, $\gamma$)$^{16}$O reaction rate. Additionally, the values of $\lambda_d$ between the considered $^{12}$C($\alpha$, $\gamma$)$^{16}$O reaction rate uncertainties are not significant. Furthermore, the depth of each TDU is independent of the $^{12}$C($\alpha$, $\gamma$)$^{16}$O reaction rate.
  
  \item The IFMR produced in this work has a similar trend to that of \citet{weidemann00}, who also consider a similar mass-loss prescription during the AGB. The IFMRs of \citet{renedo10} and \citet{salaris09} show a much steeper gradient and they consider the \citet{VW93} mass-loss prescription during the AGB.
  
  \item We find that the final ages of the sequences are in general independent of the considered reaction rate. However, during the final cooling track, we find differences up to 10\% between the adopted rates and high/low limits. This is true for both those rates taken from NACRE and \citet{an16}. This difference in the cooling time agrees with the works of \citet{prada02, salaris10, isern13}.
  
  \item The final C/O ratio in the core shows a similar trend to that at the end of the CHB. The oxygen abundance increases slightly due to the diffusion processes. The final oxygen mass fraction for NACRE\_A and NACRE\_H sequences are greater than the values derived by \citet{horowitz10} for crystallisation of a C/O core. The reaction rates from \citet{an16} agree closely with the derived values of \citet{horowitz10}. As such, future works should consider a lower reaction rate than that of NACRE when considering the crystallisation process of \citet{horowitz10}.

  \item The inner structure of the star is affected by the uncertainties within the considered reaction rates, particularly those from NACRE. The position and height of the carbon peak is significantly affected by the difference between the adopted rate and high/low limits of the reaction rate for the NACRE considerations. This may affect the modes in which pulsations can occur during the ZZ Ceti instability strip \citep{corsico06, romero12}.
  
\end{enumerate}

Although we analyse the possible evolutionary stages where more accurate $^{12}$C($\alpha$, $\gamma$)$^{16}$O reaction rates are needed, a deeper analysis of some effects are still required. For instance, a quantification of how the pulsation modes of sdB's and ZZ Ceti stars are affected, for example. Furthermore, we conclude that a lower reaction than that of NACRE\_A is favourable for the \citet{horowitz10} considerations of crystallisation, however, this must be further analysed as well. By limiting the uncertainties of $^{12}$C($\alpha$, $\gamma$)$^{16}$O reaction rates to 10\% of the adopted rate, as in \citet{an16}, reports a much better consistency of stellar parameters.

\section*{Acknowledgements}

BTP, ADR and SOK acknowledge support  by  CNPq and PRONEX-FAPERGS/CNPq. This study was financed in part by the Coordena\c{c}\~ao de Aperfei\c{c}oamento de Pessoal de N\'ivel Superior - Brasil (CAPES) - Finance Code 001. This  research  has  made  use  of NASA's Astrophysics Data System. AGI acknowledges support from the Netherlands Organisation for Scientific Research (NWO). We also thank developers of the \textsc{MESA} software, which was used extensively in this work. Finlly, we thank the anonymous referee for their input to make it a more complete work. 

\section*{Data Availability}

The data is available upon request to the corresponding author.



\bibliographystyle{mnras}
\interlinepenalty=10000
\bibliography{ref.bib}



\appendix
\section{Convection during CHB for An rates} \label{An_appendix}

\begin{figure}
    \centering
    \includegraphics[width=\linewidth]{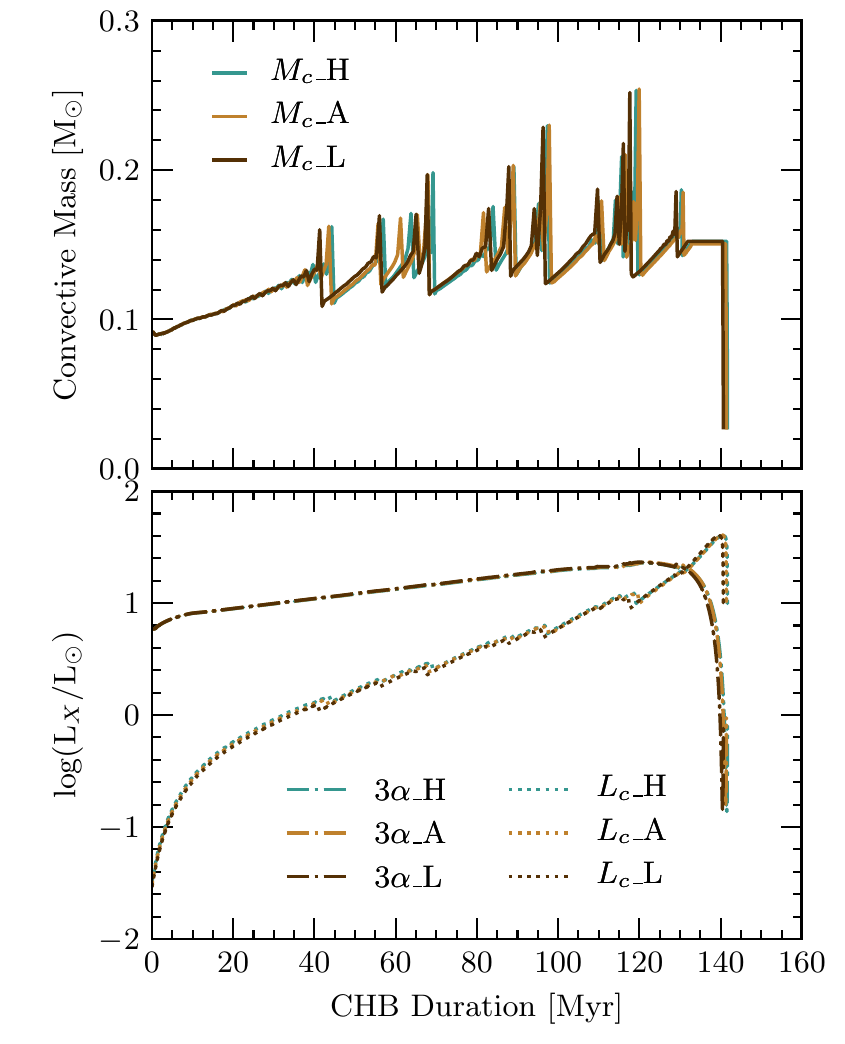}
    \caption{History of the convective mass (top panel), 3$\alpha$ luminosity and the luminosity of the $^{12}$C($\alpha$, $\gamma$)$^{16}$O reaction during the CHB (bottom panel). The history is given in terms of the CHB duration. This plot in particular considers all \citet{an16} prescriptions for the $^{12}$C($\alpha$, $\gamma$)$^{16}$O reaction rate for an initial mass of $M_i = 2.45\, M_\odot$. Blue lines represent An\_H, orange-brown depicts An\_A and dark-brown shows An\_L. Furthermore, the solid line represents the convective mass, dotted lines show the luminosity of the $^{12}$C($\alpha$, $\gamma$)$^{16}$O reaction and dot-dash lines portray the 3$\alpha$ luminosity.}
    \label{fig: ME_an}
\end{figure}

Figure~\ref{fig: ME_an} shows the CHB history of the convective mass and the luminosities of the 3$\alpha$ process and the $^{12}$C($\alpha$, $\gamma$)$^{16}$O reaction, for the reaction rates taken from \citet{an16}. This figure is analogous to that of Figure~\ref{fig: ME} which shows the same for the NACRE rates. We provide this figure to prove that we do not find any significant difference between the number of mixing episodes, luminosity from the 3$\alpha$ process and the $^{12}$C($\alpha$, $\gamma$)$^{16}$O reaction. Thus, the high/low limits for the $^{12}$C($\alpha$, $\gamma$)$^{16}$O reaction rate from \citet{an16} does not affect the CHB in terms of energy production, mixing episodes or CHB duration. This was not found for the NACRE case, which is discussed in Section~\ref{CHB}.

\section{Additional AGB Models} \label{model_appendix}

\begin{figure}
    \centering
    \includegraphics[width=\linewidth]{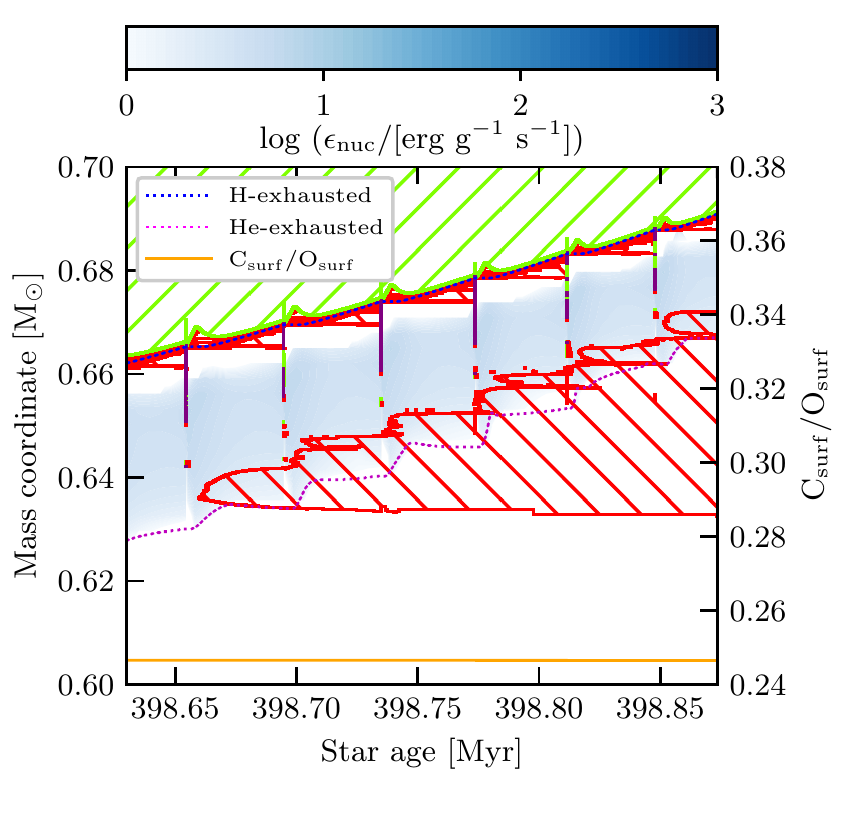}
    \caption{Kippenhahn diagram for during the TP-AGB for the case of $M_i = 3.05\, M_\odot$ of the original models. We represent the mass co-ordinate on the first y--axis and the surface C/O ratio on the second y--axis. Both values are plotted against the age of the sequence. This model did not consider convective overshooting at boundary of the He--exhausted core which inhibited the TDU. The colour bar measures the energy generation rate from nuclear reactions. The blue dotted line represents the helium core mass while the purple dotted line represents the He--exhausted core. Green slashed regions show convection and the red back slashed regions represent where regions of the star are semi-convective. Finally, purple areas are where overshooting occurs.}
    \label{fig: no_tdu}
\end{figure}

Figure~\ref{fig: no_tdu} shows the Kippenhahn diagram for the case of $M_i = 3.05\, M_\odot$ during the TP-AGB in the original NACRE\_A models. We represent the mass co-ordinate on the first y--axis and the surface C/O ratio on the second y--axis. Both values are plotted against the age of the sequence. These models did not consider convective overshooting around the border of the He--exhausted core. Green slashed areas show convective regions, red back slashed areas represent semi-convective regions and the purple regions are where overshooting occurs. The purple dotted line shows the history of the He--exhausted core mass and the blue dotted line represents the history of the helium core mass. The colour bar measures the energy generation rate from nuclear reactions. The solid orange line represents the C/O ratio at the surface. It can be seen that the overshooting occurs close to the envelope boundary and there is no overshooting about the semi-convective region of the He--exhausted core. As a result of this, we do not observe TDU episodes in the original models. We can be sure that there are no TDU episodes because of the lack of change in helium core mass and that the surface C/O ratio remains constant, which would change if TDUs were experienced \citep{frost96, herwig99, karakas02, weiss09, romero15, DG17, marigo20}.

\begin{figure}
    \centering
    \includegraphics[width=\linewidth]{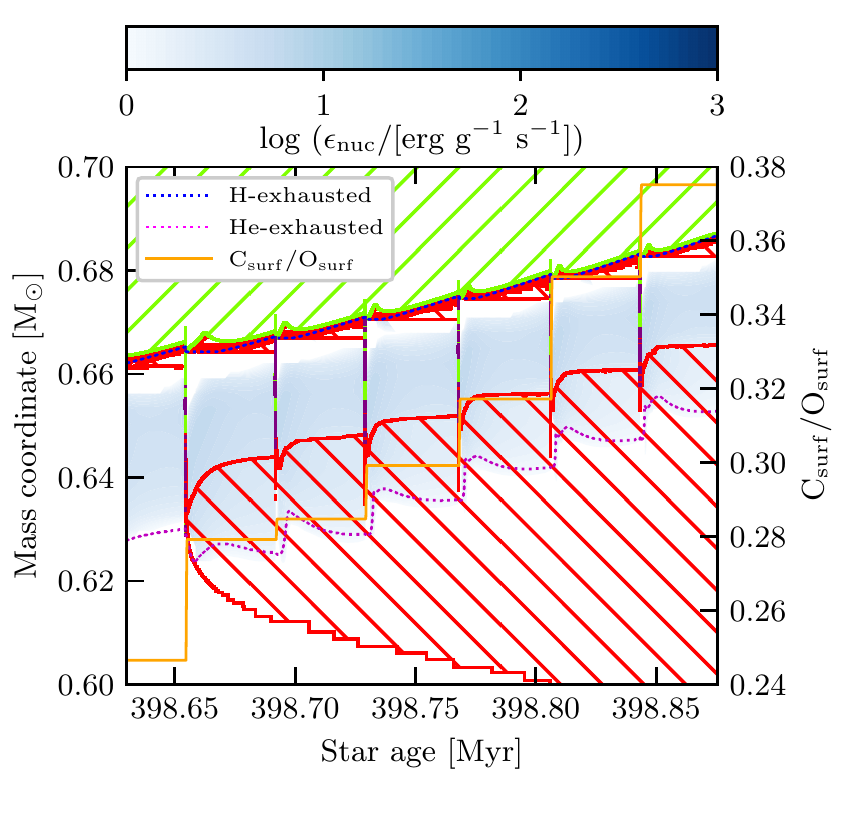}
    \caption{Kippenhahn diagram for during the TP-AGB for the case of $M_i=3.05\,M_\odot$ of the new models. We represent the mass co-ordinate on the first y--axis and the surface C/O ratio on the second y--axis. Both values are plotted against the age of the sequence. The colour bar measures the energy generation rate from nuclear reactions. This model considered convective overshooting at all convective boundaries, allowing fro TDUs to occur. The blue dotted line represents the helium core mass while the purple dotted line represents the He--exhausted core. Green slashed regions show convection and the red back slashed regions represent where regions of the star are semi-convective. Finally, purple areas are where overshooting occurs.}
    \label{fig: tdu}
\end{figure}

Figure~\ref{fig: tdu} shows the same as Figure~\ref{fig: no_tdu} but allows for convective overshooting at each boundary. We find that, with the new prescription, convection and overshooting extends throughout the helium buffer. For this reason material can be "dredged-up" from the core to the surface. This results in the helium core mass and He--exhausted core masses changing with each convective episode - an outcome of TDU episodes \citep{frost96, herwig99, karakas02, weiss09, romero15, DG17, marigo20}. Furthermore, we find an increase in the surface C/O ratio with each TDU as material travels from the stellar interior to the surface. The surface C/O ratio, however, remains less than 1. This indicates a larger overshooting parameter is required for M--star to C--star transitions.

\label{lastpage}
\end{document}